\documentclass[12pt,a4paper]{article}
 
\usepackage{graphicx} \usepackage{float}

\newcommand\BibTeX{{\rmfamily B\kern-.05em \textsc{i\kern-.025em b}\kern-.08em
T\kern-.1667em\lower.7ex\hbox{E}\kern-.125emX}}

\title{Numerical Study of Nonlinear Dynamics of a Population System with Time Delay}

\author{Ivan ~N.~Dushkov$^a$, Ivan~Jordanov$^{b,~c}$, Nikolay ~K.~Vitanov$^c$}

\date{$^{a}$ Sofia University "St. Kl. Ohridski", Bulgaria,\\ $^b$ University of National and World Economy, Sofia, Bulgaria, \\ $^c$ Institute of Mechanics, Bulgarian Academy of Sciences}

\begin{document}
\maketitle

\begin{abstract}
Mathematical models of interacting populations are often constructed 
as systems of differential equations, which describe how populations change 
with time. Below we study one such model connected to the
nonlinear dynamics of a system of populations in presence of time
delay. The consequence of the presence of the time delay is that the
nonlinear dynamics of the studied system become more rich, e.g.,
new orbits in the phase space of the system arise which are dependent on the time-delay parameters. In more detail we introduce a time delay and generalize 
the model system of differential equations for the interaction of three
populations based on generalized Volterra equations in which the growth rates and competition coefficients of populations depend on the number of members of all populations
\cite{Dimitrova2001a},\cite{Dimitrova2001b} and then numerically solve the system with and without time delay.
We use a modification of the method of Adams for the numerical solution
of the system of model equations with time delay. By appropriate selection
of the parameters and initial conditions we show the impact of the delay
time on the dynamics of the studied population system.

\end{abstract}

\textbf{Keywords}: Population dynamics; Generalized Volterra equations; Time-delay differential equations; Phase space; Adams method; Nonlinear dynamics; Chaos.

\section{Introduction}
Models based of differential and difference equations are widely used in  many areas of
natural and social sciences \cite{d1} \cite{fowler}, \cite{haber}, \cite{m1} - \cite{m3}, \cite{murray}, \cite{panchev}, \cite{nikolova1}, \cite{nikolova2}, \cite{vdk1}, \cite{vdk2}, \cite{vx1}-\cite{vx4}, \cite{zhang}. Many of these models contain
autonomous differential equations and because of this the theory of
dynamical system gained a status of  an essential tool for analysis of
dynamics of many systems, e.g. in many areas of   economic analysis
particularly since computers has become commonly available \cite{meiss},\cite{sternberg}.
We note especially the application of theory of stochastic processes and dynamics
with delay to different fields of science - fluid mechanics, economics, social sciences
population dynamics and medicine \cite{Dimitrova2015}, \cite{gopal}, \cite{demir}, \cite{pran}, \cite{v1} - \cite{vv1}.
The inclusion of time delay  into the dynamic models leads to changes in their properties \cite{kyang}, \cite{thomas}. Below we introduce time delay in the model of Dimitrova-Vitanov 
\cite{Dimitrova2000}-\cite{Dimitrova2004} and investigate the changes in the dynamics with
increasing time delay. The model of Dimitrova and Vitanov without time delay  describe competition and adaptation
in a system of populations.  Populations  compete for resources and this competition often leads to significant changes in the number of individuals of populations. Populations react to changes by a greater or lesser adaptation. Those who adapt better, have better chances of survival. Clearly, it is important to model both processes simultaneously. In the model of Volterra 
the parameters describing the rate of growth and the rate of interaction among populations are constants. But the change in the number of members of populations
leads to a change in the number of encounters between them. This can lead to a changes in the number of population members and the coefficients of the interaction. Thus the model of Dimitrova and Vitanov generalizes some classic models, such as the models of Volterra kind. 
From the point of view of mathematics the model equation in the Dimitrova-Vitanov model are
ordinary differential equations. Below we shall introduce time delay in these equations and we shall assume that the time delay parameter is independent on the time.
\section{Mathematical formulation of the problem}
The generalized equation of Lotka - Voltera is: 
\begin{equation} \label{glv}
\frac{dN_{i}(t)}{dt}=r_{i}N_{i}(t)\bigg[1-\sum_{j=1}^{n} \alpha_{ij} N_{j}(t)\bigg], \hskip .2cm i=1,2,...,n,
\end{equation}
where $N_{i}(t)$ is the density of the members of the $i-$th competing population, $r_{i}$ is ratio of increase in the density of individuals of $i-$th population,
and $\alpha_{ij}$ are the coefficients of the the interaction, showing how
$j-$th population affects the $i-$th population.
In this model, the coefficients representing the factors and interactions are constants. However, the change in the density of populations  leads to a change in the number of encounters between population members. This in turn can lead to a change of both the ratios of increase and rates of interaction. This is a manifestation of the adaptation of the corresponding population systems to changes in environmental conditions.  
Considering this adaptation Dimitrova and Vitanov \cite{Dimitrova2000}-\cite{Dimitrova2004} have proposed the following relationships for the coefficients  $r_{i}$ and $\alpha_{ij}$ of the members for each of the n competing populations:
\begin{eqnarray}\label{rgen}
	r_{i}= r_{i}^{0} \bigg[1 + \sum_{k=1}^{n} r_{ik} N_{k} + \sum_{k,l=1}^{n}
	r_{ikl} N_{k} N_{l} + 
	\sum_{k,l,m=1}^{n} r_{iklm} N_{k} N_{l} N_{m}+
	\dots \bigg],
\end{eqnarray}

\begin{eqnarray}\label{alpgen}
	\alpha_{ij}= \alpha_{ij}^{0} \bigg[1 + \sum_{k=1}^{n} \alpha_{ijk} N_{k} +
	\sum_{k,l=1}^{n} \alpha_{ijkl} N_{k}N_{l}  
	+ 
	\sum_{k,l,m=1}^{n} \alpha_{ijklm} N_{k} N_{l} N_{m}+
	\dots \bigg].
\end{eqnarray}
Further Dimitrova and Vitanov have discussed the simplest, yet most natural form of the above relations, keeping only linear members.   
Thus, the relationships for the coefficients of the growth factors and the number of the reaction of members of the systems of agents become:
\begin{equation}\label{r}
r_{i}=r_{i}^{0} \left[ 1+ \sum_{k=1}^{n} r_{ik} N_{k} \right],
\end{equation}
\begin{equation}\label{alpha}
\alpha_{ij}=\alpha_{ij}^{0} \left[ 1+ \sum_{k=1}^{n} \alpha_{ijk}N_{k} \right].
\end{equation}
Taking into account (\ref{r}) and (\ref{alpha}) the system of ordinary differential equations for modeling the dynamics of system of interacting populations in the presence of adaptation becomes:
\begin{eqnarray}\label{modeq}
	\frac{d N_{i}}{dt}= r_{i}^{0} N_{i} \{1-\sum_{j=1}^{n} [\alpha_{ij}^{0}- r_{ij}]
	N_{j}
	-\sum_{j=1}^{n} \sum_{l=1}^{n} \alpha_{ij}^{0} [\alpha_{ijl} + \nonumber \\
	+ r_{il}]
	N_{j} N_{l}
	-\sum_{j=1}^{n} \sum_{k=1}^{n} \sum_{l=1}^{n} \alpha_{ij}^{0} r_{ik}
	\alpha_{ijl} N_{j} N_{k} N_{l} \}.
\end{eqnarray}

The obtained model is a generalization of various models of dynamics of population. For an example if
\begin{eqnarray}\label{modeq1}
	r_{i}^{0}=f_{i}^{0} \nonumber\\
	r_{i}^{0}\alpha_{ij}^{0} =-f_{ij}-b_{ij}^0 
	\\
	r_{i}^{0}\alpha_{ij}^{0}(\alpha_{ijk}+r_{ik})=-b_{ijk} \nonumber \\
	r_{i}^{0}\alpha_{ij}^{0}+ r_{il}\alpha_{ijl}r_{ik}=0, \nonumber
\end{eqnarray}
then one obtains
\begin{eqnarray}\label{modeq22}
	\frac{d N_{i}}{dt}= N_{i}\bigg[f_i^0+\sum_{j=1}^{n} (\alpha_{ij}^{0}+f_{ij})N_{j}
	+\sum_{j,k=1}^{n}b_{ijk}N_{j} N_{l}\bigg].
\end{eqnarray}
Thus the model is reduced to the model discussed by Arneodo et
al.\cite{Arneodo1980}, plus additional members, accounting for the adaptation effects.
	\par 
	Let us now introduce the constant time delay $\tau > 0$ in the model. We obtain the
	following system of ordinary differential equations with time delay:
	\begin{eqnarray}\label{modeq3}
		\frac{d N_{i}(t)}{dt}= N_{i}(t-\tau)\bigg[f_i^0+\sum_{j=1}^{n}(\alpha_{ij}^{0}+f_{ij})N_{j}(t-\tau)
		+ \nonumber \\
		\sum_{j,k=1}^{n}b_{ijk}N_{j}(t-\tau) N_{l}(t-\tau)\bigg],\end{eqnarray}
	or:
	\begin{eqnarray}\label{modeq4}
		\frac{d N_{i}(t)}{dt}= r_{i}^{0} N_{i}(t-\tau)\{1-\sum_{j=1}^{n} [\alpha_{ij}^{0}- r_{ij}]
		N_{j}(t-\tau)
		- \nonumber \\\sum_{j=1}^{n} \sum_{l=1}^{n} \alpha_{ij}^{0} [\alpha_{ijl}+ 
		+ r_{il}]
		N_{j}(t-\tau) N_{l}(t-\tau)
		- \nonumber \\
		\sum_{j=1}^{n} \sum_{k=1}^{n} \sum_{l=1}^{n} \alpha_{ij}^{0} r_{ik}
		\alpha_{ijl} N_{j}(t-\tau) N_{k}(t-\tau) N_{l}(t-\tau) \}.
	\end{eqnarray}

Let us assume that the solution is sufficiently smooth  and for the right side of the equations of the system we may use the decomposition in powers of $\tau$:
\begin{equation}
\frac{d N_{i}(t)}{dt}= F_{i}(t-\tau)
\end{equation}
\begin{equation}
F_{i}(t-\tau)=F_{i}(t)-\tau \frac{d F_{i}(t)}{dt}+...+(-1)^n\frac{\tau^n}{n!} \frac{d^n F_{i}(t)}{dt^n}+\dots,
\end{equation}
where:
$$\frac{dF_{i}(t)}{dt}=\sum_{j=1}^{n} \frac {\partial F_{i}}{\partial N_j}\frac{dN_{i}(t)}{dt},$$ 
$$\frac{d^2F_{i}(t)}{dt^2}=\sum_{j,k=1}^{n} \frac {\partial^2 F_{i}}{\partial N_j\partial N_k}\frac{dN_{i}(t)}{dt}\frac{dN_{k}(t)}{dt}+\sum_{j=1}^{n} \frac {\partial F_{i}}{\partial N_j}\frac{d^2N_{i}(t)}{dt^2},$$ $$\dots$$
\par 
Then we can discuss solutions to a system (\ref{modeq4}) as:
$$N_i(t)=\sum_{\alpha=0}^{\infty}(-1)^{\alpha}\frac{\tau^{\alpha}}{\alpha !}N_{i,\alpha}(t),$$
where $N_{i,\alpha}(t)$ are functions that must be determined by the sequence of equations for the
different degrees of $\tau$ obtained from (\ref{modeq4}). For an example if $\alpha = 0$ we obtain 
the solution without time delay. Obtained sequence of equations is quite complicated for analytical solution and because of this below we shall discuss numerical solution of the model system with
time delay.
\section{Numerical solution}
We note that the presence of time delay
reduces almost to $0$ the possibility for analytical study of corresponding system of model equation. Because of this we have to solve numerically
the corresponding equations. Below we discuss briefly our approach for
numerical solution. First of all for the solution of the problem we need to set initial functions $\phi_{i}(t)$ in the range $[-\tau , 0]$. We will assume that these  functions are the solution to the system, but without any time delay.
In general the system we want to solve has the form:
\begin{equation}
\frac{dN_{i}(t)}{dt}=F_i\big[N_{1}(t-\tau),...,N_{n}(t-\tau)\big].
\end{equation}
Let us have the initial conditions $\phi_{i}(t)$ of the above kind defined in the interval $[-\tau , 0)$. We divide the interval of $k$ equal parts by the points $\{-\tau+lh\}$ where $l=0,...,k$. Because of the requirement $-\tau+kh=0$ to step $h$ we receive $h=\tau/k$. Now we can find an approximation of the solution at the point $t=mh$, $m>0$.
We integrate the above equations within the range of $t_{m-l}$ to $t_m$:
	\begin{equation}\int_{t_{m-l}}^{t_m}\dot{N_i}(t)dt=\int_{t_{m-l}}^{t_m}dN_i(t)=\int_{t_{m-l}}^{t_m}F_i\big[N_{1}(t-\tau),...,N_{n}(t-\tau)\big]dt,
	\end{equation}
	or
	\begin{equation}
	N_i^m-N_i^{m-l}=\int_{t_{m-l}}^{t_m}F_i\big[N_{1}(t-\tau),...,N_{n}(t-\tau)\big]dt.
	\end{equation}
Next we make the substitution $t=t_m+hx$. The result is:
\begin{equation}
	N_i^m-N_i^{m-l}=h\int_{-l}^{0}F_i\big[N_{1}(t_m+hx-\tau),...,N_{n}(t_m+hx-\tau)\big]dx.
\end{equation}
Taking into account that
	$$F_i\big[N_{1}(t+hx-\tau),...,N_{n}(t+hx-\tau)\big]=\frac{dN_{i}(t+hx)}{dt},$$ we obtain:
\begin{equation}
N_i^m-N_i^{m-l}=h\int_{-l}^{0}\dot{N_i}(t_m+hx)dx.
\end{equation}
	Now we  apply Newton's formula for interpolating backwards (starting with point $t$):
	\begin{eqnarray*}	\dot{N}_i(t_m+hx)=\dot{N}_i^m+\frac{x}{1!}\Delta\dot{N}_i^{m-1}+\frac{x(x+1)}{2!}\Delta^2\dot{N}_i^{m-2}+... \nonumber \\
	+\frac{x(x+1)...(x+k-1)}{k!}\Delta^k\dot{N}_i^{m-k}+R_k(x),
\end{eqnarray*}
	where finite difference forward is
	$$\Delta^l\dot{N}_i=\sum_{l=0}^{k}(-1)^{k+l} {{k}\choose{l}} \dot{N}_i^{m-k+l},$$
	The result is

	\begin{eqnarray}
		\frac{{N}_i^{m+l}-{N}_i^{m-l}}{h}=\int_{-l}^0\bigg(\dot{N}_i^m+\frac{x}{1!}\Delta\dot{N}_i^{m-1}+ \nonumber \\
\frac{x(x+1)}{2!}\Delta^2\dot{N}_i^{m-2}+\frac{x(x+1)...(x+k-1)}{k!}\Delta^k\dot{N}_i^{m-k}\bigg)dx+\int_{-l}^0R_k(x)dx.
	\end{eqnarray}
In order  to obtain the value of the solution at the point $t_m$, we must use the values in the points $t_{m-jk}$, which is why when calculating the first approximation (for point $0$), we can use the above method only when $l = 1$. In order to obtain the $s$-th  approximation we  use the method  for $l$ from $1$ to $s$. Note that for $l = 0$ the above method has a local error of approximation $O(h^{k + l})$.
We are interested in the change of the solution of the system in presence of a time delay as compared to the solution without presence of time delay. The system for the case without time delay is solved by the method of Runge-Kutta method of fourth order. 
\par
Let us now apply the above methodology. We are going to show results for a particular case of the discussed model system, namely for the system of three populations ($i=1,2,3$)
\begin{eqnarray}\label {modeq2}
	\frac{d N_{i}}{dt}= N_{i}\bigg[f_i^0+\sum_{j=1}^{3} (\alpha_{ij}^{0}+f_{ij})N_{j}
	+\sum_{j,k=1}^{3}b_{ijk}N_{j} N_{l}\bigg],
\end{eqnarray}
	
	$$b_{11}^0=-k_1, b_{12}^0-k_1, b_{13}^0=-k_2, b_{21}^0=k_1, b_{22}^0=k_1,$$ $$ b_{23}^0=-k_2, b_{31}^0=-k_3, b_{32}^0=-k_2, b_{33}^0=-k_3,$$
	$$b_{ijk}=b, f_{ij}=f,$$
	$$f_1^0=2k_1+k_2, f_2^0=-k_1, f_3^0=2k_2+k_3.$$
	The fixed points of this model system are as follows \cite{Dimitrova2004}. The first fixed pointis
	$$N_1=\frac{-R_1(k_1+k2)-3k_1^2+2k_1k_2+k_2^2}{2k_1(k_2-k1)}, N_2=\frac{R_1}{k_2-k_1}, N_3=0$$
	where $R_1$ is the real root of the equation:
	$$bZ^2+(2k_1^2-6bk_1-2bk_2)Z+(2k_1^3+2k_1^2k_2+2fk_1k_2+bk_2^2+6fk_1^2+9bk_1^2+6bk_1k_2).$$
	The second fixed point is:
	$$N_1=N_2=1, N_3=R_2,$$
	where $R_2$ is the real root of the equation:
	$$bZ^2+(4b+f-k_2)Z+2f+4b+k_2.$$
If a time delay is presented then the system (\ref{modeq2}) is transformed to a system of the king (\ref{modeq3}). Some effects of presence of the time delay in the model equations may be
illustrated as follows.  First we consider the case without the presence of a time delay and choose the parameter values as follows: $k_1 = 0.5$, $k_2 = 0.1$, $k_3 = 1.05$, $f = 10^{-6}$, $b = 10^{-5}$. In this case 
the attractor in the phase space is  a limit cycle. Let's include the time lag and gradually start to increase 
the value of the delay parameter. For small values of the time lag $\tau$ there is no qualitative change in the phase trajectory of the system. With increasing value of $\tau$  the  behavior in phase space  is transformed  from limit cycle to a chaotic attrcator. For values of  $\tau$  larger than $0.5$ the solution is no more confined, which is  related to reaching the limits of the stability of the numerical algorithm. 
\par 
The presence of time delay may lead to a transition in the opposite direction: from chaotic to cyclic behaviour. In order to demonstrate this we choose the following values of parameters: $k_1 = 0.5$, $k_2 = 0.1$, $k_3 = 1.43$, $f = 10^{-6}$, $b = 10^{-5}$. In the case without delay the model system of equations  describes   chaotic behaviour in the phase space. For small values of the time lag the behaviour in the phase spaces remains chaotic, i.e., small value of the time lag is not sufficient to destabilize the chaotic attractor of the system. By increasing the value of the time lag the behavior in the phase space changes from chaotic to periodic one. When values of the time lag $\tau$ are approximately $0.01$  a limit cycle appears that remains stable with increasing values of the time lag up to $\tau = 0.5$. 
\section{Concluding Remarks}
In this paper, we discuss a nonlinear model of the dynamics of interacting populations with time delay.  The models of kind (\ref{modeq3}) and (\ref{modeq4}) may have many applications as they take into an account not only the current state of the studied system but also account for the past states of the system. In addition these models
account for the influence of environment on the growth rates as well as
for the influence of the environment on the coefficients of the
interaction among the studied populations. Because of this  models like the model described above are very flexible and  may be applied for description of phenomena in other areas such as  biology,  economy, technological evolution social sciences, etc.
The presence of time delay complicates the model equation and what can be done is to perform numerical studies of the dynamics of corresponding system.
We have shown that the method of Adams can be applied successfully for numerical solution of the model system of ordinary differential equations in the presence of a time delay. On the basis of the numerical studies one can draw several conclusions about the impact of the time lag on the behavior of the system.
First of all and  as it can be expected for small values of the time lag there is no  substantial influence of the time lag on the kind of the attractor
of the system in phase space. But the
increasing values of the delay parameter may lead to such qualitative 
change in the behaviour. 
This change  may be associated with both the transition from periodic to chaotic attractor and a transition from chaotic attractor to a limit cycle.

\section*{Acknowledgements} 
This work contains results, which are supported
by the UNWE project for scientific researchers with grant agreement No.
NID NI - 21/2016.



\end{document}